\documentclass[amsmath,amssymb,aps,twocolumn,prb,longbibliography,floatfix,superscriptaddress]{revtex4-2}
\usepackage{graphicx,bm,times}
\usepackage[utf8]{inputenc}
\usepackage{braket}
\usepackage{amsfonts}
\usepackage{color}

\usepackage[breaklinks]{hyperref}
\hypersetup{
    pdfstartview={FitH},
    colorlinks=true,
    linkcolor=blue,
    citecolor=blue,
    urlcolor=blue
}

\begin{document}

\title{Local electron correlation effects on the Fermiology of the weak itinerant ferromagnet ZrZn$_2$}
\author{Wenhan~Chen}
\affiliation{H.~H.~Wills Physics Laboratory,
University of Bristol, Tyndall Avenue, Bristol, BS8 1TL, United Kingdom}

\author{A.~D.~N.~James}
\affiliation{H.~H.~Wills Physics Laboratory,
University of Bristol, Tyndall Avenue, Bristol, BS8 1TL, United Kingdom}

\author{S.~B.~Dugdale}
\affiliation{H.~H.~Wills Physics Laboratory,
University of Bristol, Tyndall Avenue, Bristol, BS8 1TL, United Kingdom}

\date{\today}

\begin{abstract}
The Fermi surface topology plays an important role in the macroscopic properties of metals. It can be particularly sensitive to electron correlation, which appears to be
especially significant for the weak itinerant ferromagnet ZrZn$_{2}$.
Here, we look at the differences in the predicted Fermi surface sheets of this metallic compound in its paramagnetic phase for both density functional theory (DFT) and the combination of DFT with dynamical mean field theory (DFT+DMFT). The theoretical spectral functions evaluated at the Fermi level were used along with calculations of the electron-positron momentum density (also known as the two-photon momenutm density) in $k$-space to provide insights into the origin of certain features of the Fermi surface topology. We compare this two photon momentum density  to that extracted from the positron annihilation experimental data (Phys.~Rev.~Lett.~92,~107003~(2004)). The DFT+DMFT densities are in better agreement with the experiment than the DFT, particularly with regard to the flat bands around the $L$ and $W$ high symmetry points. The experimental neck around $L$, which relates to a van Hove singularity, is present in DFT+DMFT but not in the DFT. We find that these flat bands, and as such the Fermi surface topology, are sensitive to the many body electron correlation description, and show that the positron annihilation technique is able to probe this. This description is significant for the observed behavior such as the Lifshiftz transition around the quantum critical point.
\end{abstract}

\maketitle

\section{Introduction}
Weak itinerant ferromagnets are a class of materials with small ordered magnetic moments ($M_O$) and low Curie temperatures ($T_C$) \cite{morosan2017}. ZrZn$_{2}$, which has a C15 cubic Laves phase structure with lattice parameter of ${7.393}$~\AA, is a rare example of a member of this class ($T_C \approx 28$~K, $M_O \approx 0.17 \mu_B$ per formula unit) \cite{PhysRevB.72.184436} and one which has been shown to host a variety of phenomena including pressure-tuned quantum criticality and non-Fermi liquid behavior both close to and away from the quantum critical point \cite{Grosche1995,PhysRevB.72.184436,Kabeya2013,Smith2008}. Interest in this material was significantly piqued by an apparent coexistence of superconductivity and ferromagnetism in high-quality samples (with very low residual resistivities of the order of $0.62 \mu\Omega {\rm cm}$) \cite{Pfleiderer2001}.
Such coexistence, predicted decades ago by Ginzburg ~\cite{ginzburg1957antagonie}, was thought to be due to triplet pairing through spin fluctuations~\cite{PhysRevB.22.3173}. On the other hand, the Fulde-Ferrell-Larkin-Ovchinnikov (FFLO) state ~\cite{larkin1969quasiclassical,PhysRev.135.A550}, influenced by the Fermiology, was another candidate to explain this coexistence ~\cite{PhysRevLett.88.187004,Major2004}. P-wave superconductivity was also predicted to exist in the paramagnetic phase and it was thought that the reason no superconductivity had been observed was due to sample disorder~\cite{Powell_2003}. However, subsequent investigations by Yelland {\it et al.} found no evidence of a bulk superconducting state in ZrZn$_2$~\cite{PhysRevB.72.184436}, and in fact the previously measured superconductivity was due to a superconducting surface layer induced by spark erosion~\cite{PhysRevB.72.214523}.

Yamaji {\it et al.} pointed out that the metamagnetic transitions in ZrZn$_2$ around the so-called marginal quantum critical point, accompanied by a discontinuous change in the Fermi surface topology (a Lifshitz transition), belong to an unconventional class of symmetry-breaking transition ~\cite{yamaji2007quantum}. Previous measurements of the magnetic, transport, and thermodynamic properties strongly support this point ~\cite{kabeya2012non,PhysRevB.85.035118}. Density functional theory (DFT) calculations show flat bands at the Fermi level around the $L$ and $W$ high symmetry points which generate a van Hove singularity responsible for the weak ferromagnetic ordering in the material and play an important role in the metamagnetic transition in a weak magnetic field, Lifshitz transitions, and non-Fermi-liquid behavior ~\cite{IGOSHEV2022128107}. The resulting sensitivity of the predicted Fermi surface to details of the calculation can be seen in the literature, as discussed below.
An early DFT study indicated that a small Fermi surface pocket is created by the very flat band at $L$~\cite{PhysRevB.37.3489}. While later results by Singh and Mazin~\cite{PhysRevLett.88.187004} gave a similar Fermi surface topology with both studies predicting a connected `pillow' Fermi surface sheet around $W$ (described by Singh and Mazin as a `pancake'), there were differences. Singh and Mazin did not observe this small Fermi surface pocket at $L$. The Lifshitz transition in this material is tightly connected with the van Hove singularities around $L$ and $W$.

There have been measurements of the ferromagnetic Fermi surface near the quantum critical point at temperatures as low as 30mK using the quantum oscillatory de Haas-van Alphen effect by Yates {\it et al.}~\cite{Yates2003}. In order to investigate the paramagnetic Fermi surface topology away from the quantum critical point, Major {\it et al.}~\cite{Major2004} measured the Fermi surface of ZrZn$_2$ by positron annihilation in the paramagnetic phase (at $T \approx 80$K). Interestingly, the positron annihilation experiment on paramagnetic ZrZn$_2$ revealed the existence of an extra Fermi surface neck  around the $L$ point which was not seen in those previous calculations \cite{PhysRevLett.88.187004,PhysRevB.37.3489}. Note that the positron annihilation results were compared to calculations made using the linear-muffin-tin-orbital (LMTO) method which actually showed a neck at the $L$ point. However, since those LMTO calculations were made in the atomic-sphere-approximation (ASA), it could be expected that the full-potential calculations would more accurately describe the bands and Fermi surface.
This difference strongly indicates that the flat band around $L$ point may in fact be (partially) below the Fermi level as indicated by a recent DFT+dynamical mean field theory (DMFT) study~\cite{PhysRevB.102.085101}.

In positron annihilation experiments, the positron annihilates with the electron in the material producing two $\gamma$-photons.  By measuring the deviation from anti-collinearity it is possible to infer two components of the electron-positron momentum distribution, which is also called the two-photon momentum distribution (TPMD). Note that the full three-dimensional density can be recovered using tomographic techniques applied to data measured with different crystallographic orientations (see for example Ref.~\cite{Kontrym_Sznajd-Cormack_1990}). The TPMD will include the effects of electron-positron correlations which could potentially obscure the signatures of many-body electron correlations. However, it has been shown that the measured electron-positron momentum distribution is actually strongly influenced by the electron correlations in the material and is not hidden by the electron-positron correlations~\cite{etde_20717331}. This may imply that many significant differences between the predictions from DFT and positron annihilation experiments may in fact be due to the limitations of the approximation to the exchange-correlation functional used in the DFT framework for electrons. Previously, positron annihilation and DFT+DMFT has been used to provide insight into the Hubbard $U$ value of Ni~\cite{ce.we.16} and electron correlation effects in V~\cite{Weber_2017}. It is therefore necessary to include the electron correlations described beyond the DFT framework in calculations of the electron-positron momentum distribution.

A recent DFT+DMFT study has highlighted that the DMFT Fermi-liquid-like electron self-energies subtly change the Fermi surface topology of ZrZn$_2$ \cite{PhysRevB.102.085101}. This includes the prediction of the existence of a Fermi surface neck which touches the Brillouin zone at $L$, in agreement with the positron measurements. DMFT includes electron correlations beyond that approximated within DFT by including the many-body local dynamical electron interactions~\cite{ge.ko.96}. These interactions can prove vital in modeling strongly correlated systems where these local electron correlations originating in the more localized (e.g. $d$ and $f$) orbitals, make a significant contribution and give rise to phases such as the Mott insulator~\cite{ge.ko.96,held2007electronic}. For ZrZn$_2$, the inclusion of the local electron correlations in the partially occupied Zr $4d$-states around the Fermi level has been shown to have a significant influence on the important features of the electronic structure and Fermi surface topology which is dominated by the flat bands around the $L$, $W$, and $K$ points~\cite{PhysRevB.102.085101}.

In this paper, we perform DFT and DFT+DMFT calculations which are compared against the positron annihilation data which was previously measured by Major {\it et al.}~\cite{Major2004}. We compare these theoretical predictions with this experimental data to highlight the importance of electron correlations included in DMFT on the Fermi surface topology. We find that these sheets are mainly influenced by the changes to the flat bands around $L$, $W$, and $K$ located around the Fermi level indicating the significant impact the local electron correlations have on the macroscopic properties, considering that it is these flat bands which influence the Lifshitz transition in this material~\cite{yamaji2007quantum}.

\section{Methods}
\subsection{DFT+DMFT Calculations}
In this study, we have used the full potential APW+lo {\sc elk} code~\cite{elk} in combination with the toolbox for research on interacting quantum systems (TRIQS) library~\cite{triqs}. This so-called {\sc elk}+TRIQS DFT+DMFT framework is described in Ref.~\cite{james_2021} together with further discussion of interfacing the APW+lo DFT basis with the DMFT Anderson's impurity model found in Ref.~\cite{aichhorn2009}. The ZrZn$_{2}$ DFT calculation was converged on a $32 \times 32 \times 32 $ Monkhorst-Pack $k$-mesh of 897 irreducible $k$-points in the first Brillouin zone. The Perdew-Burke-Ernzerhof (PBE) generalized gradient approximation (GGA) functional~\cite{PhysRevLett.77.3865} was used to approximate the exchange-correlation interactions within the DFT framework. The DFT outputs were interfaced to the TRIQS library by constructing Wannier projectors, as described in Ref.~\cite{james_2021}, for all the Zr $4d$-states within a correlated energy window of $[-8,10]$~eV which itself spans the occupied Zn $3d$ and partially filled Zr $4d$ bands; this energy window captures the equivalent states used in  Ref.~\cite{PhysRevB.102.085101}. The paramagnetic DMFT calculation was implemented using the continuous time quantum Monte Carlo (CT-QMC) solver within the TRIQS/CTHYB application~\cite{seth2016} with the Slater interaction Hamiltonian parameterized by the Hubbard interaction $U=2.5$~eV and Hund exchange interaction $J=0.3$~eV, as used in Ref.~\cite{PhysRevB.102.085101}.
We approximated the double counting in the fully localized limit, as implemented in Ref.~\cite{PhysRevB.102.085101}.
We used the fully-charge self-consistent (FCSC) DFT+DMFT method with an inverse temperature $\beta =$ 40 eV$^{-1}$ (\mbox{$\sim$290 K}) and $8.96\times10^{8}$ Monte Carlo sweeps for each cycle of the impurity solver. The spectral functions were calculated by analytically continuing the DMFT self-energy using the {\em LineFitAnalyzer} technique of the maximum entropy analytic continuation method implemented within the TRIQS/Maxent application \cite{PhysRevB.96.155128}.

\subsection{DFT+DMFT Two Photon Momentum Density Implementation}
The two dimensional angular correlation of annihilation radiation (2D-ACAR) experiments used positrons from a $^{22}$Na radioisotope source, as described  by Major {\it et al.}~\cite{Major2004}. A
thermal positron will annihilate with an electron within the sample which subsequently results in the emission of two $\gamma$-photons which in their rest frame will be directed antiparallel to each other to conserve momentum. In the laboratory frame, however, the $\gamma$-photons deviate from being anti-parallel mainly due to the momentum of the annihilated electron, whose momentum is typically significantly larger than that of the thermalized positron. Since the angular deviation is proportional to momentum, the measured deviation gives a direct insight into the electron momentum distribution. The $\gamma$-photons are recorded by two position-sensitive detectors located either side of the sample chamber, giving two momentum components. The third component of momentum, which is expressed as a Doppler shift in the photon energies, cannot be measured with the same resolution.
Thus the measured spectra is a projection of the TPMD along the axis through the sample chamber connecting the centre of the two detectors.

The connection between the (real space) wave functions and the TPMD can be simply understood via the Fourier transformation~\cite{PhysRevB.71.233103} (assuming spin degeneracy) as

\begin{equation}
\begin{aligned}
  \rho^{2\gamma}(\mathbf{p}) &= \sum_{\mathbf{k}, \eta}n_{\mathbf{k}, \eta}\left|\int\limits_{V} \sqrt{\gamma(\mathbf{r})}\psi_{\mathbf{k}, \eta}^{e}(\mathbf{r})\psi^{p}(\mathbf{r}) \exp (-i \mathbf{p} \cdot \mathbf{r})  \mathrm{d} \mathbf{r}\right|^{2}\\
  &= \sum_{\eta}n_{\mathbf{k}, \eta}\left|C_{\mathbf{k}+\mathbf{G},\eta}\right|^{2},
\end{aligned}
\label{TPMD}
\end{equation}
where $\psi^{p}(\mathbf{r})$ is the positron wave function and $\psi_{\mathbf{k}, \eta}^{e}(\mathbf{r})$ is the Bloch electron wave function with eigenstate index $\eta$ at point $\mathbf{k}$ in the $k$-mesh used. The $\gamma(\mathbf{r})$ quantity is the electron-positron enhancement factor which is designed to describe the electron-positron correlations. If this factor is assumed to be a constant, then this is referred to as the independent particle model. However,  different approximations can be adopted to improve agreement between experiment, such as the Drummond enhancement~\cite{drummond_2010,drummond_2011} which was used in this study. The last line of Eq.~\ref{TPMD} introduces $C_{\mathbf{k}+\mathbf{G},\eta}$ which is a shorthand for the Fourier coefficients of the combined enhanced electron-positron wave function product.

To calculate the DFT+DMFT TPMD, we followed a similar process as that used to determine the DFT+DMFT electron momentum density (EMD) as described by James {\it et al.}~\cite{james2020magnetic} which itself is based on the DFT EMD computation methodology implemented by Ernsting {\it et al.}~\cite{Ernsting_2014}. A new set of electron wave functions and occupation numbers are determined by diagonalizing the total DFT+DMFT density matrix, derived from the DFT+DMFT lattice Green's function by summing over the Matsubara frequencies, and which is generally non-diagonal in the Kohn-Sham basis.  Further details can be found in Refs.~\cite{james_2021,james2020magnetic}. The resulting orthonormal Lowdin-type basis electron wave functions and diagonal occupation numbers can be then directly used in Eq.~\ref{TPMD} as it is only the electron wave functions which have been updated to include the many-body electron-electron interactions. Both approximations used to generate $\gamma(\mathbf{r})$ and $\psi^{p}(\mathbf{r})$ are unchanged with respect to the DFT calculation, although these quantities will likely change between the DFT and DFT+DMFT calculations as they are dependent on the electron density which is updated in each FCSC DFT+DMFT cycle. It should be noted that this method benefits from not requiring any analytical continuation of the DFT+DMFT quantities. For the DFT and DFT+DMFT TPMDs calculations here, we used a denser Monkhorst-Pack $k$-mesh of $60 \times 60 \times 60 $ which equates to 5216 irreducible $k$-points in the first Brillouin zone, and a maximum momentum cut-off of 8.5 a.u. at which the TPMD has converged to zero.

\subsection{The Lock-Crisp-West Theorem}
Although the signatures of Fermi surfaces are of course present in the `real momentum'-space ($p$-space) TPMD, developing a full understanding of the Fermi surfaces through a direct analysis of the TPMD is not easy due to the absence of (discrete) translational invariance. However, the TPMD in $p$-space, can be transformed into crystal momentum space ($k$-space) by translating contributions to the TPMD back into the first Brillouin zone through the reciprocal lattice vectors, $\mathbf{G}$. This new reduced-TPMD (or RTPMD) can be thought of as the occupation number for the electrons \textit{as seen by the positron} for each $k$-point summed over the bands. In such a distribution, the Fermi surfaces are now much clearer. The transformation of TPMD to the RTPMD (i.e., from $p$-space to $k$-space) is based on the so-called Lock-Crisp-West theorem (LCW)~\cite{Lock_1973}.
As discussed above, and with reference to Eq.~\ref{TPMD}, one can directly express the RTPMD, $\rho^{2\gamma}(\mathbf{k})$, by summing over the reciprocal lattice vectors $\mathbf{G}$ by using
\begin{equation}
  \rho^{2\gamma}(\mathbf{k})=\sum_{\eta} \sum_{\mathbf{G}}n_{\mathbf{k}, \eta}\delta(\mathbf{p}-(\mathbf{G}+\mathbf{k}))\left|C_{\mathbf{k}+\mathbf{G},\eta}\right|^2.
  \label{eqlcw1}
\end{equation}
While the summation over the reciprocal vectors $\mathbf{G}$ in Eq. \ref{eqlcw1} is in principle infinite, in practice this can be truncated at some momentum cut-off since the TPMD rapidly goes to zero. This is because the wave function of the positively charged positron doesn't overlap much with the most tightly bound (core) electrons which would contribute at the highest momentum (localization in real space, means delocalization in momentum space). The delta function ensures conservation of momentum.

The Fermi surface can be identified as the boundary between occupied and unoccupied states in $k$-space. In practice, the locations of the Fermi surfaces are determined by first finding the maximum in the magnitude of the derivatives of the RTPMD. The Fermi surfaces are then contoured in the RTPMD at the corresponding isovalues (evaluated at those maxima). In order to facilitate the comparison between experiment and calculation (including evaluation on particular planes and along particular directions), and with the benefits of better computational resources which permit this, the tomographic reconstruction of the 3D density from the deconvoluted measured projections has been repeated \cite{Dugdale_1994,Fretwell_1995,Kontrym_Sznajd-Cormack_1990,Major2004,Samsel-Czekala2003}.

In the reconstructed experimental RTPMD described above, the related experimental statistical noise $\sigma\left[\rho(\mathbf{k})\right]$ was quantified following by that implemented by Samsel et al.~\cite{Samsel-Czekala2003} within $k$-space. A set of $M$ independent reconstructions were performed on theoretical TPMDs to which (different) random noise was added. The standard error $\sigma$ in $\rho(\mathbf{k})$ at a particular $k$-point was evaluated as:
\begin{equation}
  \sigma\left[\rho(\mathbf{k})\right]=\left[\frac{1}{M}\sum\limits_{i=1}\limits^{M}[\rho_{i}(\mathbf{k})-\overline{\rho}(\mathbf{k})]^{2}\right]^{\frac{1}{2}},
  \label{eqerrb}
\end{equation}
where $\rho_{i}(\mathbf{k})$ is the $i^{\rm th}$  reconstruction and $\overline{\rho}(\mathbf{k})$ is the average reconstruction.  We subsequently used $\sigma\left[\rho(\mathbf{k})\right]$ to represent the experimental uncertainty in the usual chi-squared `goodness-of-fit' parameter to help quantify how well the DFT and DFT+DMFT agree with the experimental data.

\section{Results}
\subsection{Spectral Function}
\begin{figure}[t!]
 \centerline{\includegraphics[width=\linewidth]{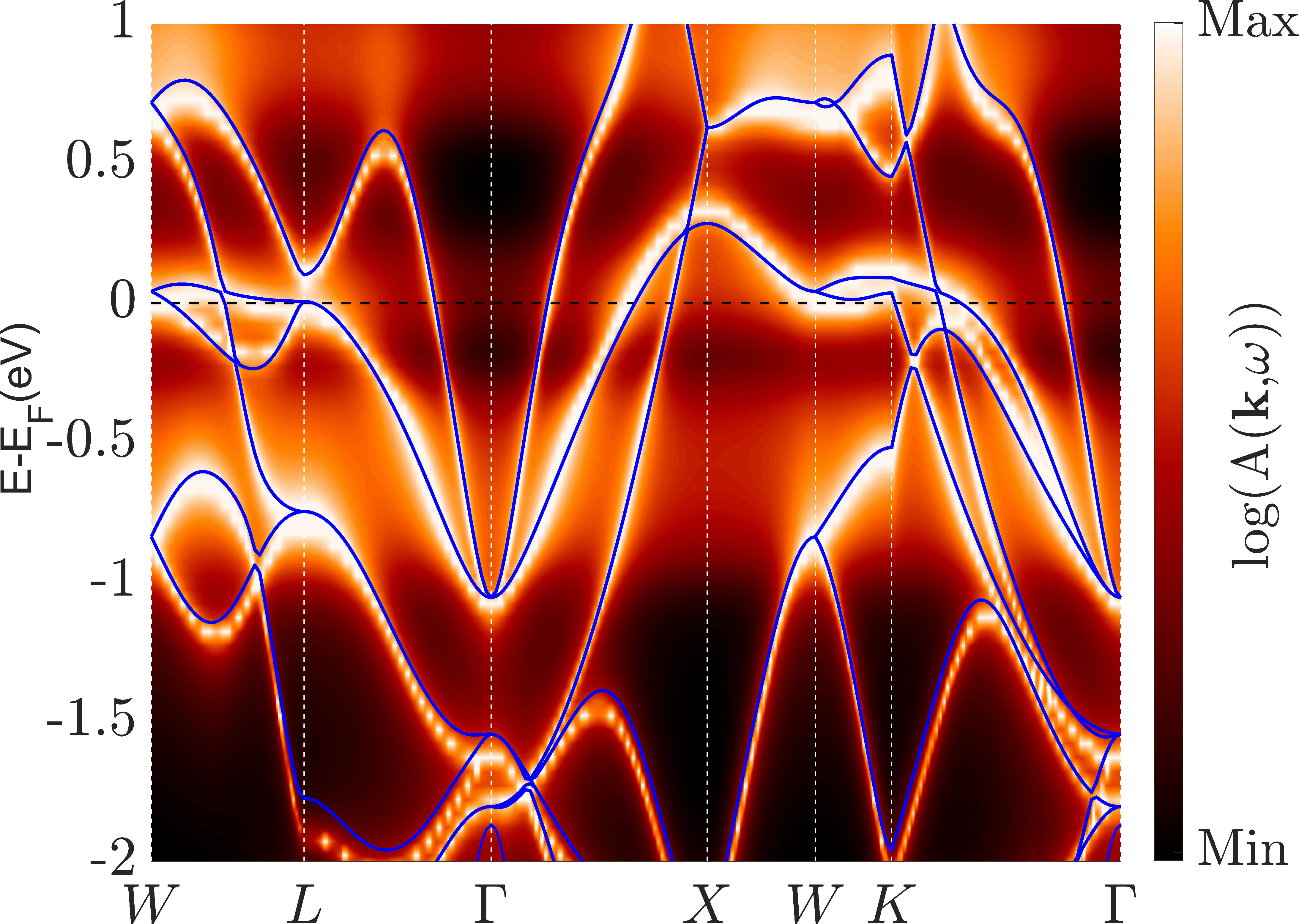}}
 \caption{Electronic structure of paramagnetic ZrZn$_{2}$ revealed through the band structure from DFT (blue solid lines) and the DFT+DMFT spectral functions (shown as a color map). To illustrate the spectral functions clearly, the logarithm  ($\log \left(A \left(\mathbf{k},\omega\right)\right)$) is plotted. Note the differences between DFT and DMFT, particularly along the $X$---$W$---$K$ and $W$---$L$ segments, where the DMFT spectral weight grazes the Fermi level. }
\label{dmftband}
\end{figure}

Figure~\ref{dmftband} shows the DFT bandstructure overlayed onto the logarithm of the DFT+DMFT spectral function. The inclusion of electron correlations from DMFT clearly introduces broadening to the spectral function, as expected, along with visible changes to the positions of the band centers.
The most significant change in the DFT+DMFT spectral function with respect to the DFT bandstucture is in the bands along the $X$---$W$---$K$ and $W$---$L$. Along these paths it can be seen that the flat bands cross the Fermi level at different points in the Brillouin zone. The bands which dispersed just above the Fermi level (just grazing it) along these paths in DFT have now, with the inclusion of the electron correlations in the DMFT, been shifted downwards so that they cross the Fermi level, resulting in the emergence of new features in the Fermi surface. This is in good agreement with Ref.~\cite{PhysRevB.102.085101} with a few exceptions described below. The logarithm of DFT+DMFT $k$-resolved spectral function in Fig.~\ref{dmftband} shows weakly $k$-dependent incoherent weight around $\sim$ --0.6 and 0.8 eV and the DFT band just above the Fermi level at $W$ is shifted down in energy in the DFT+DMFT calculation so that this grazing band which in DFT was just above the Fermi level along $W$---$K$ is now lying partially below the Fermi level in the DFT+DMFT calculation. We expect this to be the consequence of the interaction Hamiltonian as well as the lower temperature used in the DMFT part of our calculations. In our calculation, the full interaction Hamiltonian has been used which includes the spin-flip and pair-hopping interactions which the density-density interaction Hamiltonian used in Ref.~\cite{PhysRevB.102.085101} omits. We note that the magnitude of the imaginary part of the real frequency-dependent self-energy at $\omega$~=~0 is small so that the predicted electronic structure does not deviate too much from the Fermi-liquid regime at the temperature used for the DMFT cycle; this is similar to the DFT+DMFT results in Ref.~\cite{PhysRevB.102.085101}.

\begin{figure}[t!]
 \centerline{\includegraphics[width=0.85\linewidth]{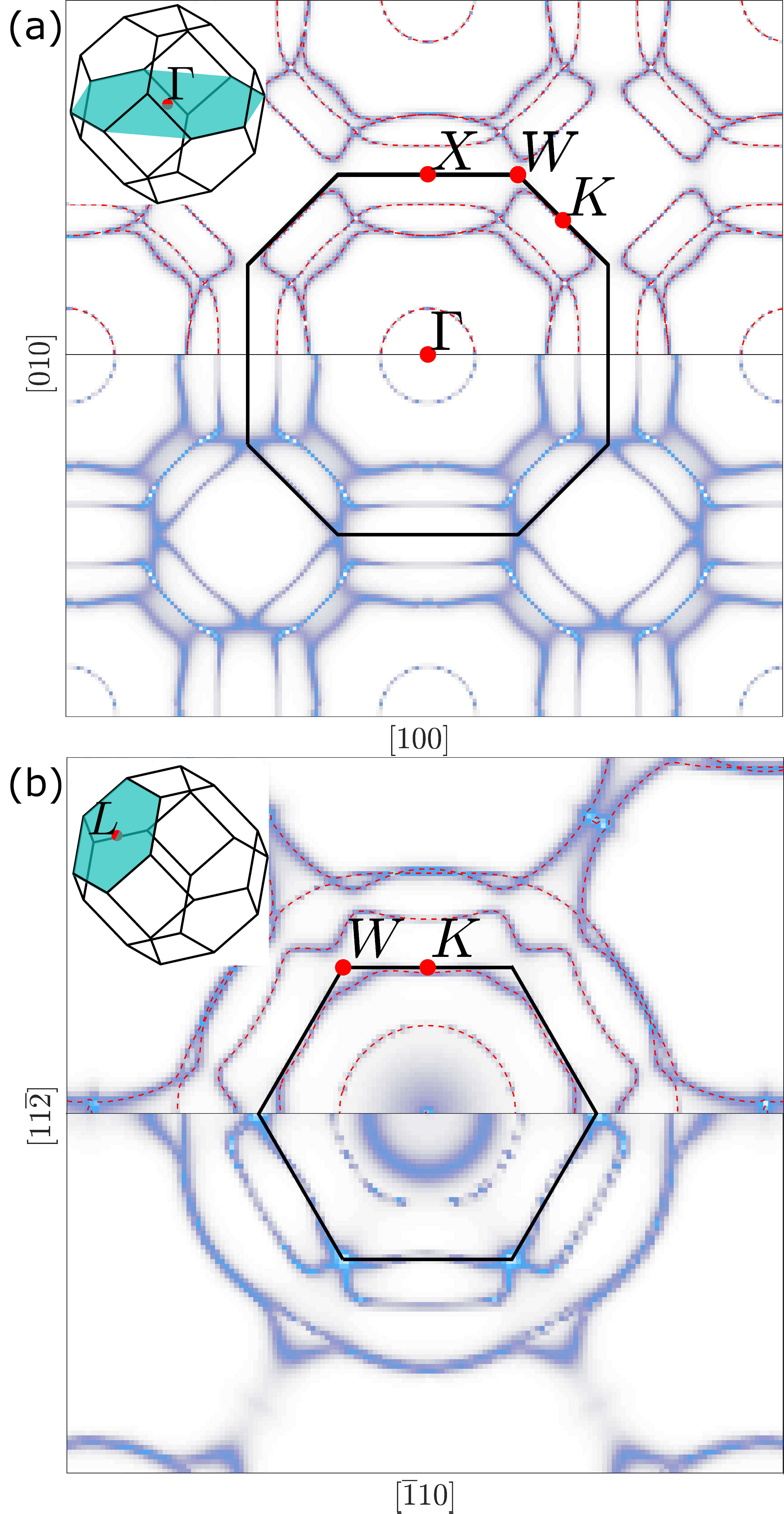}}
 \caption{Contour of spectral functions at $\omega=E_{F}$ on (a) the plane with $k_z=0$ centered at $\Gamma$ point and (b) the plane that perpendicular to reciprocal vector $(111)$ and centered at $L$. The upper panel of each figure is computed from the DFT calculation. The colormap of the upper panel is the DFT spectral functions calculated with a small artificial imaginary energy of 0.01 eV. The red dashed lines are the DFT Fermi surfaces on those planes. The colormaps of the bottom half of each figure are computed from the DFT+DMFT calculation. The location of each plane is schematically shown at the top left corner of the sub-figure.}
\label{fs2d}
\end{figure}

\begin{figure*}[t!]
    \centerline{\includegraphics[width=0.95\linewidth]{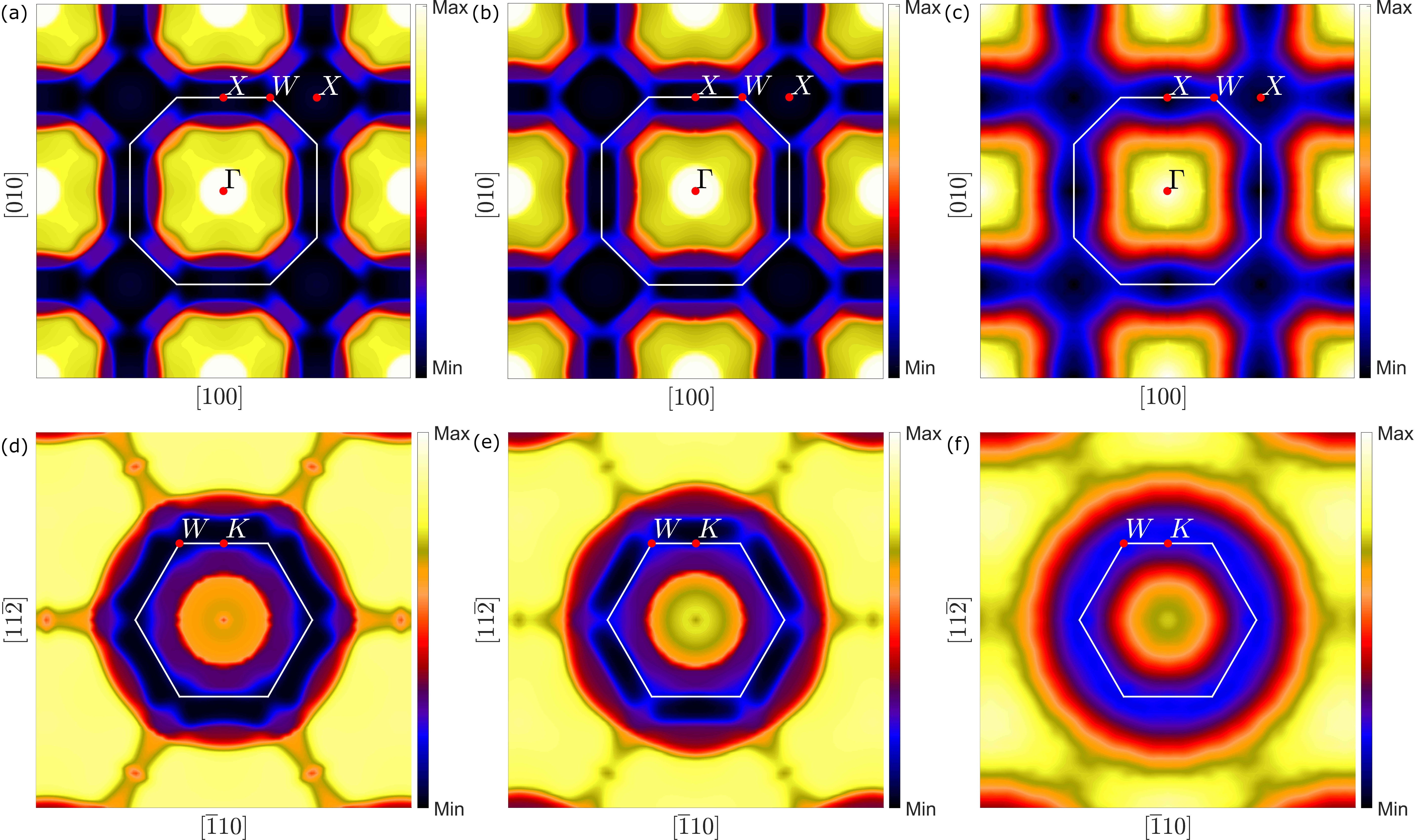}}
    \caption{Slices through the RTPMD calculation on (a-c) the $(001)$ plane through $\Gamma$ and (d-f) the $(111)$ plane through $L$ (see Fig.~\ref{fs2d}) from the DFT calculation (a and d), the DFT+DMFT calculation (b and e) and the experiment (c and f). The colorbar shows the normalized RTPMD for each subfigure.}
\label{tpmdjoin}
\end{figure*}

The DFT and DFT+DMFT spectral functions in Fig.~\ref{fs2d}~(a) and (b) are evaluated at the Fermi level and calculated on the two dimensional (2D) planes shown by the corresponding insets of Fig.~\ref{fs2d}~(a) and (b). When both Figs.~\ref{fs2d}~(a) and (b) are taken in conjunction with the DFT band structure and DFT+DMFT spectral function in Fig.~\ref{dmftband}, we see that there are clear changes going from DFT to DFT+DMFT in the spectral weight present at the Fermi level which here is evidence of the changes in the Fermi surface topology within these planes. Most of the significant changes in the 2D spectral functions occur in the regions around $W$, $K$, and $L$, where the previously unoccupied DFT flat bands just above the Fermi level relating to the van Hove singularity have changed in DFT+DMFT,  now crossing the Fermi level as can be seen in Fig.~\ref{dmftband} along $W$---$L$ and $W$---$K$. Both Figs.~\ref{fs2d}~(a) and (b) show the appearance of DFT+DMFT spectral weight at $W$ which indicates that there is now a portion of the corresponding Fermi surface sheet there. In the DFT+DMFT 2D spectral function shown in Fig.~\ref{fs2d}~(b), there now exists additional spectral weight around $L$ (with respect to the DFT 2D spectral function) which relates to the formation of a large neck in the Fermi surface topology, in agreement with Ref.~\cite{PhysRevB.102.085101}. Note that in both Figs.~\ref{fs2d}~(a) and (b) the DFT Fermi surface (i.e., the Fermi surface from the DFT energy eigenvalues) in those planes is plotted on top of the DFT 2D spectral function. In Fig.~\ref{fs2d}~(b), there is indeed a very small DFT Fermi surface at $L$ which is in fact a small hole pocket. It is only just visible and appears as a dot, being smaller than that seen in Ref.~\cite{PhysRevB.37.3489}; this small pocket can also be inferred from the DFT band structure near $L$ along the path of $W$---$L$ in Fig.~\ref{dmftband}. Interestingly, this Fermi surface sheet appears to be quite sensitive to how the electron correlations are modeled considering that the Fermi surface topology changes from a hole pocket at $L$ in DFT to an electron neck enclosing $L$ in the DFT+DMFT results.

\begin{figure*}[t!]
    \centerline{\includegraphics[width=0.92\linewidth]{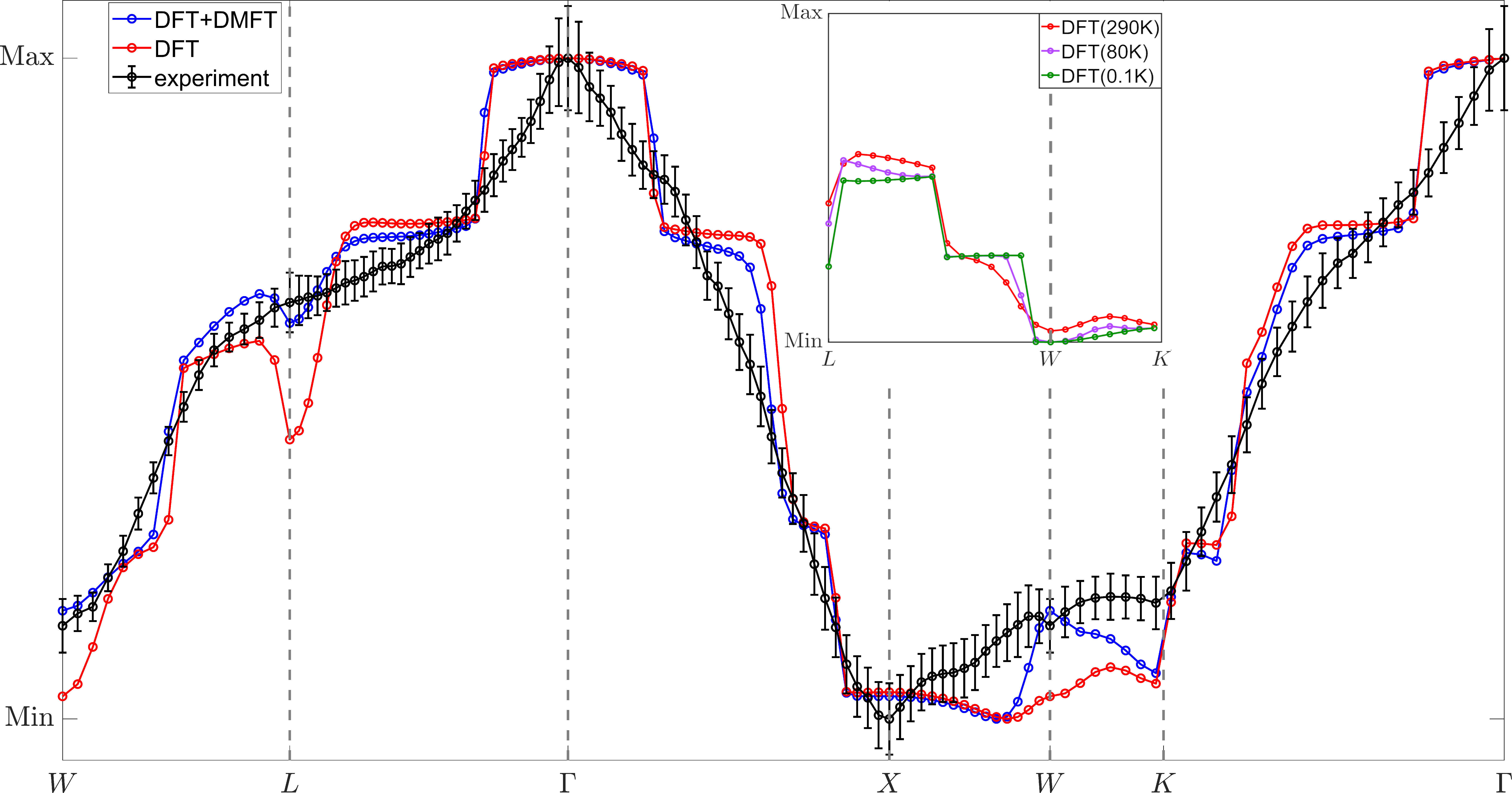}}
    \caption{The RTPMD plotted along the same high symmetry path that was used in the band structure plot (Fig.~\ref{dmftband}). The normalized RTPMD are shown with solid lines guiding the eye between the data point which are shown as circles. The experimental errors were determined as described by Eq.~\ref{eqerrb} in the main text. The theoretical RTPMDs used a temperature of $\sim$ 290 K, and the inset shows the temperature dependence of the DFT RTPMDs along the high symmetry paths where there are the flat bands around the Fermi level.}
\label{tpmdbdpath}
\end{figure*}

\begin{figure}[t!]
    \centerline{\includegraphics[width=0.95\linewidth]{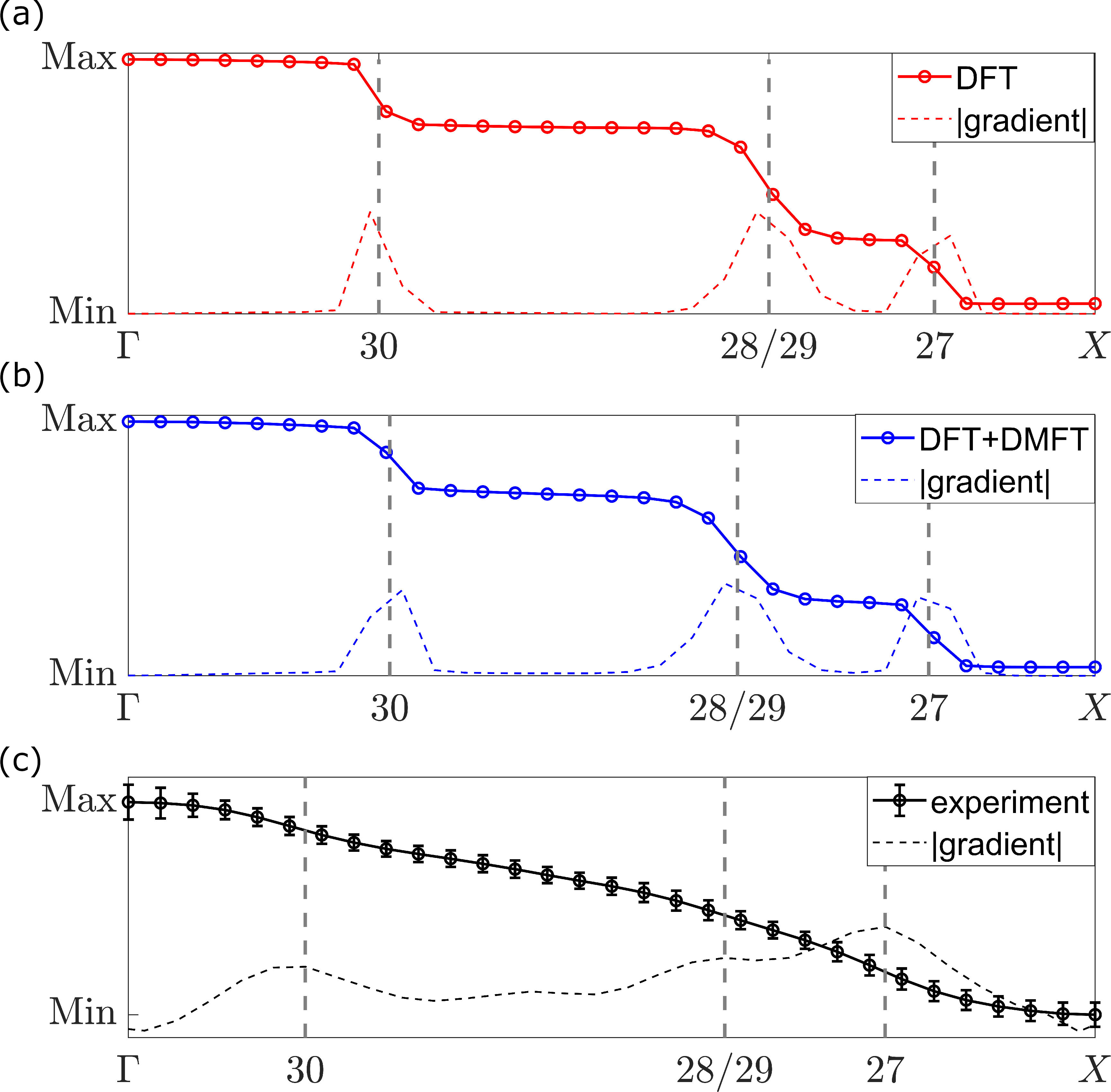}}
    \caption{The RTPMD of ZrZn$_{2}$ along the $\Gamma$---$X$ direction for (a) the DFT calculation, (b) the DFT+DMFT calculation, and (c) experiment.  The experimental errors along $\Gamma$---$X$ were determined as described by Eq.~\ref{eqerrb} in the main text. The solid lines show the TPMD while the absolute value of the gradient is shown with dashed lines. The numbers (27---30) show the corresponding band indices, as defined in Ref.~\cite{Major2004}, as discussed in the main text.}
\label{tpmdgxg}
\end{figure}

\begin{figure*}[t!]
    \centerline{\includegraphics[width=0.95\linewidth]{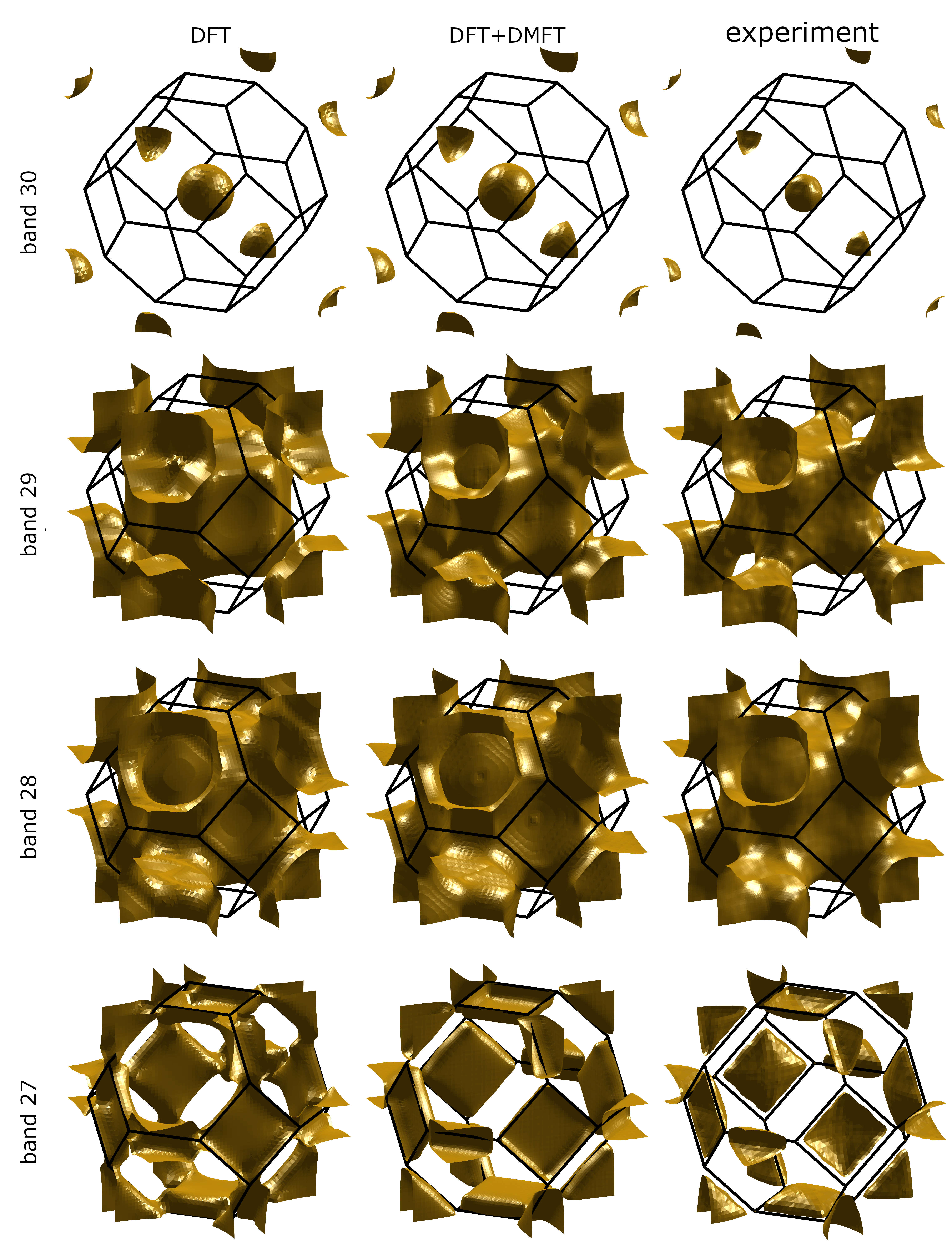}}
    \caption{The four Fermi surface sheets of paramagnetic ZrZn$_{2}$ derived from the RTPMD at the maximum values of the gradient of the RTPMD for DFT (left), DFT+DMFT (middle), and experiment (right). The band indices (27---30) refer to those in Ref.~\cite{Major2004}, as discussed in the main text.}
\label{tpmdfs}
\end{figure*}

\subsection{Reduced Two-Photon Momentum Density}

The Fermi surface can be identified by the loci of wavevectors which correspond to sharp, discontinuous changes in occupation within the Brillouin zone for a non-interacting system at zero temperature. In practice the Fermi surface can be loci of the wave-vectors at which the absolute value of the gradient of the occupation function is maximum.
Of course, at finite temperature and in the presence of electron correlations, the corresponding change in electron occupation will be smeared. However, the Fermi surface can still be identified. Here, we focus our attention on the calculated RTPMDs which can be directly compared to the corresponding experimental RTPMDs derived from the data first reported by Major \textit{et al.}~\cite{Major2004}. Before inspecting the specific changes in the RTPMD between DFT and DFT+DMFT, we note that the irreducible Brillouin zone chi-squared value from DFT+DMFT is very slightly improved (7.4) compared to the corresponding DFT value of 8.1, although both indicate that there are still significant differences. However, this shows that the inclusion of the local electron correlations from DMFT are needed for paramagnetic ZrZn$_2$. Figure~\ref{tpmdjoin} shows the RTPMD evaluated on the same two 2D planes shown in the insets of Figs.~\ref{fs2d}~(a) and (b). As one would expect, the changes which occur in the RTPMD between DFT and  DFT+DMFT reflect those seen in the spectral functions on those 2D planes in Fig.~\ref{fs2d}. The DFT+DMFT RTPMDs are more smeared than the corresponding DFT ones, owing to the additional electron correlations.
The most noteworthy changes are seen around $W$ in both planes. From the RTPMD along $X$---$W$ in Figs.~\ref{tpmdjoin}~(a-c), we see the consequence of the DFT+DMFT quasiparticle band center crossing the Fermi level around $W$ in the $\Gamma$ centered plane. Here, we see that the DFT+DMFT RTPMD is in better agreement with the experimental one at $W$, which is also clearly seen in Fig.~\ref{tpmdbdpath} in which the RTPMD is plotted along the same high symmetry path as the band structure in Fig.~\ref{dmftband}. Even though the DFT+DMFT RTPMD is in very good agreement with the experimental one at $W$, it would appear that the location of the Fermi surface associated with the quasiparticle band needs to be further away from $W$ along $X$---$W$ in order to agree with the experiment along $X$---$W$. Consequently, this would therefore likely cause the global minimum of the RTPMD to be at $X$ as in the experiment instead of that predicted in the theoretical RTPMDs. It should be noted that the distribution from $X$---$W$ is also sensitive to the positron wave function as it differs to the `pure' electron occupations along this path, but is relatively insensitive to the electron-positron enhancement model used.
On the other hand, looking at the RTPMD around $W$ in the $L$ centered plane in Figs.~\ref{tpmdjoin}~(d-f), again the DFT+DMFT RTPMD better matches the experimental RTPMD from the 2D-ACAR experiment~\cite{Major2004} which means that the quasiparticle flat band around the Fermi level from $W$---$K$ (see Fig.~\ref{dmftband}) does indeed cross the Fermi level. However, both the DFT and DFT+DMFT predicts a suppression of the RTPMD (i.e., the darker regions of Figs.~\ref{tpmdjoin}~(d) and (e)) around the edge of the first Brillouin zone, which is not seen in the experiment. These darker regions relate to the presence of the pillow Fermi surface sheet intersecting this plane when in reality, this is not the case, as seen in the experimental data in Fig.~\ref{tpmdjoin}~(f); this means that the pillow sheet is smaller in size. Regarding the distribution along $W$---$K$ shown in Fig.~\ref{tpmdbdpath}, the DFT RTPMD has a similar shape to the experimental one although the magnitude is smaller. On the contrary, the DFT+DMFT RTPMD differs by reducing in magnitude from $W$---$K$ which leads to the presence of the pillow Fermi surface sheet crossing here. Although the magnitude of the RTPMD from DFT+DMFT better agrees with the experimental one than that of DFT along $W$---$K$, the discrepancy between DFT+DMFT and experiment increases along this path to $K$. This suggests that the quasiparticle flat band around the Fermi level needs to be fully occupied along the path of $W$---$K$ which the experimental RTPMD seems to suggest to be the case. It should be noted that there is an additional contribution to the RTPMD distribution from $W$---$K$ from the positron wave functions like that seen along $X$---$W$. Also, there is a small bump in both theoretical RTPMDs at approximately midway between $W$---$K$ which is interestingly due to temperature smearing of the electron occupation, as can be seen in the temperature dependence of the DFT RTPMD distributions shown in the inset of Fig.~\ref{tpmdbdpath}. For the DFT+DMFT RTPMD, there is an additional smearing contribution from the electron correlations included  in the DMFT.

It can be seen in Fig.~\ref{tpmdbdpath} that the DFT+DMFT RTPMD is in better overall agreement with the experimental compared to the DFT. Note that the chi-squared calculated along this path for the DFT+DMFT calculation is half that of the DFT calculation. This is mostly likely because the changes in the Fermi surface are concentrated around the face of the Brillouin zone, making it particularly sensitive to changes along a path which has segments on the boundary. Both theoretical RTPMDs predict a dip at $L$ with the DFT+DMFT being shallower than the DFT. Interestingly, the dip at $L$ in the DFT+DMFT is not associated with a  Fermi surface hole pocket as one might expect. Indeed, we see an upward trend in the RTPMD along $W$---$L$ just before the dip in both the DFT and DFT+DMFT. This upward trend is due to the temperature smearing of the RTPMD, as shown in the inset of Fig.~\ref{tpmdbdpath}, from the DFT band (DFT+DMFT quasiparticle band) close to the Fermi level as well as there being an additional smearing contribution from the electron correlations in the DFT+DMFT RTPMD. The dip at $L$ in the DFT+DMFT RTPMD is likely due to the spectral weight `spilling' across the Fermi level and as such this spillage would consequently reduce the contribution to the RTPMD from the degenerate quasiparticle bands at $L$ (see Fig.~\ref{dmftband}). This spectral weight effect on the electron occupation density has been observed before in momentum densities~\cite{PhysRevLett.124.046402}. Reducing the temperature will likely suppress the hump in the DFT+DMFT just before the dip at $L$ along $W$---$L$, similar to that seen in the DFT RTPMD temperature results shown in the inset of Fig.~\ref{tpmdbdpath}. Unfortunately, lower temperature DFT+DMFT calculations are more computationally expensive and may suffer from the sign problem~\cite{RevModPhys.83.349}, and as such the DFT+DMFT RTPMD temperature dependence has not been investigated here.

There are also notable changes in the DFT+DMFT RTPMD along the other paths in Fig.~\ref{tpmdbdpath} which brings it closer in agreement with the experiment. We highlight the RTPMDs along $\Gamma$---$X$ in Fig.~\ref{tpmdgxg} which nicely shows that there is additional smearing in the DFT+DMFT RTPMD with respect to the DFT which is consistent with effect of the additional local many-body correlations. Also in Fig.~\ref{tpmdgxg}, we show the gradients of the RTPMDs and their maximums agree well with the Fermi surface locations (represent by the vertical dashes between $\Gamma$---$X$ in Fig.~\ref{tpmdgxg}) as predicted by the spectral functions (or eigenvalues in the DFT) within the resolution of the $k$-mesh. The highlighted band numbers in Fig.~\ref{tpmdgxg} correspond to the band nomenclature of Ref.~\cite{Major2004}. Here, the DFT+DMFT is in closer agreement with the experiment for bands 27, 28, and 29, but there is still a notable discrepancy with the experimental Fermi surface. On the other hand, the band 30 DFT+DMFT electron Fermi surface sheet size is slightly worse in agreement with the corresponding smaller experimental sheet compared to the DFT. Even with this being the case, the overall agreement with the experimental improves with the inclusion of DMFT.

Finally, we show each theoretical and experimental Fermi surface sheet extracted from the RTPMDs in Fig.~\ref{tpmdfs}. Here, all the previously discussed features and changes can be seen in each three dimensional Fermi surface sheet. We note that certain regions in the calculated Fermi surface sheets have distinct bumpy features on them which is the consequence of the $k$-position of the relatively sharp steps in the RTPMD. This would likely be resolved for a denser $k$-mesh which would produce a better mesh resolution but this would be more computationally demanding and have a greater $k$-mesh resolution than that possible for the experimental data. The theoretical band 30 Fermi surface sheet from DFT and DFT+DMFT is clearly too large with respect to the experiment. The emergence of the Fermi surface neck at $L$ which is associated to band 29 is observed for the DFT+DMFT and experiments, although the experimental neck is smaller than the DFT+DMFT one. The DFT, on the other hand, predicts no neck and the tiny hole Fermi surface pocket sheet seen previously is barely visible here. The topology of the Fermi surface for band 28 is very similar between the theories and experiment, but the experimental Fermi surface is slightly smaller. Lastly, the band 27 `pillow' Fermi sheet at $W$ in DFT are connected whereas in the DFT+DMFT, these pillow Fermi surface sheets are disconnected which agrees with the experimental sheets, however, these experimental `pillow' sheets are smaller than the predicted ones. This reflects the RTPMD along $W$---$K$ shown in both Figs.~\ref{tpmdjoin} and \ref{tpmdbdpath} along with the previous discussions about them.
Owing to the improved resources available for the new reconstruction, the experimental pillow sheets are in fact smaller in size, disconnected, and located around the $X$ centered plane at the Brillouin zone boundary. This disconnecting part at $W$ between each of the experimental pillow Fermi surface sheets is highly sensitive to the chosen RTPMD Fermi surface isovalue, which likely indicates that the band is only just below the Fermi surface. It is clear that these Fermi surface sheets are sensitive to the flat bands around the Fermi level near $L$, $W$, and $K$ which themselves are sensitive to the electron correlation description. Clearly the positron annihilation data is sensitive to the presence of the many-body electron correlations which are not smeared out by the electron-positron correlation contributions.

\section{Conclusions}

In describing the electronic structure of ZrZn$_2$, the inclusion of the local electron correlations from DMFT are vital as they vastly improve the Fermi surface topology resulting in superior agreement with the 2D-ACAR experimental data \cite{Major2004}. In order to analyse and compare the features of the Fermi surface from both DFT and DFT+DMFT, we used both the 2D spectral function (and DFT 2D Fermi surface) and the RTPMDs which are in excellent agreement with each other as expected. Indeed, the inclusion of DMFT captures certain features of the Fermi surface topology, such as the band 29 Fermi surface sheet neck around $L$ which is in fact a hole pocket in DFT. It is this neck feature which relates to the previously noted van Hove singularity at $L$. With the improved agreement from the DFT+DMFT calculations it appears that this feature is dependent on the local electron correlations of the Zr 4d electrons. The Fermi surface sheets derived from the RTPMD show an overall improved agreement with the experimental for the DFT+DMFT compared to the DFT. Along the $W$---$K$ path in the Brillouin zone, the DFT+DMFT RTPMD has better agreement with the experimental one, resulting in the `pillow' Fermi surface sheet to become disconnected which is also seen in the experimental data. Clearly, the flat bands around $L$, $W$, and $K$ are sensitive to the description of the many-body electron correlations which consequently affects the Fermi surface topology. There is also a clear temperature dependence in the calculated RTPMDs owing to the temperature smearing of the occupation function of the flat bands in close vicinity to the Fermi level near these high symmetry points. Finally, the positron annihilation technique proves to be a powerful tool to probe the many-body electron correlation effects considering that these are not obscured by the electron-positron correlation contributions.

\section{Acknowledgements}
Wenhan~Chen acknowledges the funding and
support from the Chinese Scholarship Council (CSC), Grant No. 201908060087. A.~D.~N.~J. also acknowledges the Doctoral Prize Fellow funding and support from the Engineering and Physical Sciences Research
Council (EPSRC).

\end{document}